\documentclass[prd,preprintnumbers,nofootinbib,aps]{revtex4}
\usepackage{graphicx}
\usepackage{dcolumn}
\usepackage{bm}
\usepackage{epsfig,amsmath}
\usepackage[usenames,dvipsnames]{color}
\usepackage[pagebackref=false, colorlinks=true]{hyperref}
\usepackage{appendix}
\definecolor{redish}{rgb}{0.7,0.2,0.0}  
\definecolor{bluish}{rgb}{0.2,0.5,0.8}

\hypersetup{linkcolor=redish,          
                  citecolor=blue,        
                  filecolor=magenta,      
                  urlcolor=bluish}          

\DeclareFontFamily{U}{rsfs}{}         
\DeclareFontShape{U}{rsfs}{m}{n}{<5> rsfs5 <6><7> rsfs7          %
  <8><9><10><10.95><12><14.4><17.28><20.74><24.88> rsfs10}{}     %
\DeclareMathAlphabet{\mathfs}{U}{rsfs}{m}{n}                     %

\newcommand{\ba}{\nopagebreak[3]\begin{eqnarray}}
\newcommand{\ea}{\end{eqnarray}}

\newcommand{\f}{\frac}

\def \d{\Delta}
\def \x{\Xi}

\def \l{\Lambda}
\def \t{\theta}
\def \O{\Omega}

\def \p{\partial}

\begin{document}

\title{Lense-Thirring Precession in Pleba\'nski-Demia\'nski spacetimes}
\author{Chandrachur Chakraborty}
\email{chandrachur.chakraborty@saha.ac.in}
\affiliation{Saha Institute of Nuclear Physics, Kolkata 700064, India}

\author{Partha Pratim Pradhan}
\email{pppradhan77@gmail.com}
\affiliation{Vivekananda Satabarshiki Mahavidyalaya,
Paschim Medinipur, WB 721513, India.}

\begin{abstract}
An exact expression of Lense-Thirring precession rate is
derived for non-extremal and extremal Pleba\'nski-Demia\'nski spacetimes.
This formula is used to find the exact Lense-Thirring precession rate in various
axisymmetric spacetimes, like: Kerr, Kerr-Newman, Kerr-de Sitter etc.
We also show, if the  Kerr parameter vanishes in 
Pleba\'nski-Demia\'nski(PD) spacetime, 
the Lense-Thirring precession does not vanish due to 
the existence of NUT charge.
To derive the LT precession rate in extremal Pleba\'nski-Demia\'nski
we first derive the general extremal condition for PD
spacetimes. This general result could be applied 
to get the extremal limit in any
stationary and axisymmetric spacetimes.

\end{abstract}

\maketitle

\section{Introduction}
The axisymmetric vacuum solutions of the Einstein
equations are used to describe the various characteristics 
of different spacetimes. The most important and physical spacetime is 
Kerr spacetime\cite{rk}, describes the rotating black hole which
possesses a finite angular momentum $J$. 
The Kerr spacetime with a finite charge $Q$ is expressed
as Kerr-Newman black hole. 
Actually, these all spacetimes 
may include the Cosmological constant $(\Lambda)$ by which 
term the complexity arises in calculations.
Without this particular constant the spacetimes
possess the two horizons, we mean, event horizon and 
Cauchy horizon. But, the presence of the cosmological constant  
leads to possess an extra horizon - the cosmological 
horizon. All these spacetimes can be taken as the special
cases of the most general axisymmetric spacetime of
Petrov type D which is first given by
Pleba\'nski and Demia\'nski (PD)\cite{pd}. This spacetime contains 
the seven parameters - acceleration,
mass $(M)$, Kerr parameter ($a$, angular momentum per unit mass),
electric charge $(Q_c)$, magnetic charge $(Q_m)$ 
NUT parameter $(n)$ and Cosmological constant.
At present, this is the most general axially symmetric
vacuum solution of Einstein field equation. It is well-known 
to us that any axisymmetric and stationary spacetime
with angular momentum (rotation) are known
to exhibit an effect called
Lense-Thirring (LT) precession whereby 
locally inertial frames are dragged
along the rotating spacetime, making 
any test gyroscope in such spacetimes
{\it precess} with a certain frequency called the LT precession
frequency \cite{lt}.
More generally, we can say that
frame-dragging effect is the property of all stationary 
spacetimes which may or may not be axisymmetric\cite{ns}.
We have also discussed this special feature in detail 
in our previous paper\cite{cm}. In that paper,
we have showed that only Kerr parameter is not 
responsible for the LT precession, NUT parameter 
is equally important to continue the frame-dragging effect
\cite{cm}. It has been shown by Hackmann and L\"ammerzahl
that LT precession vanishes (eqn.45 of \cite{hl}) 
in PD spacetimes (with vanishing acceleration of 
the gravitating source), if the Kerr parameter $a=0$.
 But, it is not the actual case. As we have shown our
previous paper, the same thing is also happened
in the PD spacetimes with {\it zero angular momentum $(J=a=0)$},
the LT precession does not vanish due to the 
presence of NUT charge $n$ ({\it angular momentum monopole}\cite{sen1})
\\ 

 Our aim is to derive the exact LT precession rates in 
non-extremal and extremal Pleba\'nski-Demia\'nski
spacetimes without invoking the weak field approximation.
So, we organize the paper as follows. In section II,
we review the general Lense-Thirring precession 
formula in stationary and axisymmetric spacetimes
and derive the exact Lense-Thirring precession rate
in Pleba\'nski-Demia\'nski(PD) spacetimes with vanishing
acceleration of the gravitating source and
discuss the exact LT precession rates in some
other stationary and axisymmetric spacetimes as the 
special cases of PD spacetimes. If the Kerr 
parameter vanishes in PD spacetimes, 
the frame-dragging effect does not vanish due
to the existence of NUT charge. It is shown in a 
subsections of sec II, as a special case of LT
precession in PD spacetimes.
In sec. III, we derive the more general extremal 
condition for PD spacetimes and discuss about 
the exact LT precession rates in PD spacetimes and 
also other various
extremal axisymmetric spacetimes as the special 
cases of PD spacetimes. A short discussion closes 
the paper which is in sec.IV.

\section{Non-extremal case}

\subsection{Pleba\'nski-Demia\'nski (PD) spacetimes}
The PD spacetime is the most general axially
symmetric vacuum solution of Einstein
equation, at present.
The line element of six parameters (as we take acceleration vanishes) PD spacetimes
can be written as (taking, $G=c=1$)\cite{hl},
\begin{eqnarray}
ds^2=-\f{\d}{p^2}(dt-Ad\phi)^2+\f{p^2}{\d}dr^2
+\f{p^2}{\x} d\t^2+\f{\x}{p^2}\sin^2\t(adt-Bd\phi)^2
\label{lnelmnt}
\end{eqnarray}

where,
\begin{eqnarray} \nonumber
p^2&=&r^2+(n-a\cos\t)^2,
\\
A &=&a \sin^2\t+2n \cos\t, B=r^2+a^2+n^2 \nonumber
\\
\d&=&(r^2+a^2-n^2)(1-\f{1}{\ell^2}(r^2+3n^2)) 
-2Mr+Q_c^2+Q_m^2-\f{4n^2r^2}{\ell^2}  \nonumber
\\
\x&=&1+\f{a^2\cos^2\t}{\ell^2} -\f{4an\cos\t}{\ell^2}
\label{defn}
\end{eqnarray}
$\f{1}{\ell^2}=\l$ denotes the Cosmological constant divided by three, 
represents the  Pleba\'nski-Demia\'nski-de-Sitter (PD-dS) spacetimes
and if $\ell^2$ is replaced by $-\ell^2$, 
it represents the PD-AdS spacetimes.
So, our metric $g_{\mu\nu}$ is thus following
\begin{eqnarray}
 g_{\mu\nu}&=&\left(\begin{array}{llll}
 -\f{1}{p^2}(\d-a^2\x \sin^2\t) & 0 & 0 & \f{1}{p^2}(A\d-aB\x \sin^2\t)\\
           0 & \f{p^2}{\d} & 0 & 0\\
           0 & 0 & \f{p^2}{\x} & 0\\
         \f{1}{p^2}(A\d-aB\x \sin^2\t) & 0 & 0 & \f{1}{p^2}(-A^2\d+B^2\x \sin^2\t)
                 \end{array}
\right)
\label{metric}
\end{eqnarray}
In our previous paper\cite{cm}, we have
already discussed about the exact expression 
of LT precession rate\cite{ns}. 
This is applicable for any non-accelerating 
stationary spacetimes. In this paper, we are going
to derive the exact LT precession rate for non-accelerating  
PD spacetimes and also some others
axially symmetric spacetimes like this. 
The expression for LT precession rate
in non-accelerating, stationary and
axisymmetric spacetime can be written as,
\begin{eqnarray}
 \O_{LT}=\f{1}{2\sqrt {-g}} \left[\left(g_{0\phi,r}-
\f{g_{0\phi}}{g_{00}} g_{00,r}\right)\p_{\theta}
- \left(g_{0\phi,\theta}-
\f{g_{0\phi}}{g_{00}} g_{00,\theta}\right)\p_r\right]
\label{lt}
\end{eqnarray}
which is same as eqn.(12) of \cite{cm}. The various metric components can be read off
from above metric (\ref{metric}). Likewise,
\begin{equation}
 \sqrt{-g}=p^2 \sin\t
\end{equation}
Substituting the metric components into eqn.(\ref{lt}) we can easily get
the LT precession rate in PD spacetimes. But, there is a 
problem in that formulation as the precession formula 
is in co-ordinate basis.
So, we should transform the precession
frequency formula from the coordinate basis to the orthonormal
`Copernican' basis: first note that
 \begin{eqnarray}
\O_{LT}&=& \O^{\theta} \p_{\theta}+\O^r \p_r \\
\O_{LT}^2&=&g_{rr}(\O^r)^2+g_{\theta\theta}(\O^{\theta})^2
\end{eqnarray}

Next, in the orthonormal `Copernican' basis at rest in the rotating
spacetime, with our choice of polar cooridnates, $\O_{LT}$ can be written as
\begin{equation}
 \vec{\O}_{LT}=\sqrt{g_{rr}}\O^r \hat{r}+\sqrt{g_{\theta\theta}}\O^{\theta} \hat{\t}
\end{equation}
where, $\hat{\t}$ is the unit vector along the direction
$\theta$. Our final result of LT precession in 
non-accelerating PD spacetime is then,

\begin{eqnarray}\nonumber
\vec{\O}_{LT}^{PD}&=&\f{\sqrt{\d}}{p}\left[\f{a(\x \cos\t+
(2n-a \cos\t)\f{a}{\ell^2} \sin^2\t)}{\d-a^2\x \sin^2\t}
-\f{a \cos\t-n}{p^2}\right]\hat{r}
\\
&+&\f{\sqrt{\x}}{p}a \sin\t \left[\f{r-M-\f{r}{\ell^2}(a^2+2r^2+6n^2)}{\d-a^2\x \sin^2\t}
-\f{r}{p^2}\right]\hat{\t}
\label{pd}
\end{eqnarray}

Now, from the above expression we can 
easily derive the LT precession rates for various
axisymmetric stationary spacetimes 
as the special cases of PD spacetime.

\subsection{Special cases}
{\bf (a) Schwarzschild and Schwarzschild-de-Sitter spacetimes :}
As the Schwarzschild and Schwarzschild-de-Sitter spacetimes both
are static and $a=\l=Q_c=Q_m=n=0$, the inertial frames are not dragged
along it. So, we can't see any LT effect in these spacetimes.
This is very well-known feature of static spacetime.
\\

{\bf (b) Kerr spacetimes :}
LT precession rate for non-extremal Kerr spacetimes 
is already discussed in detail in our
previous paper\cite{cm}. Setting $\l=Q_c=Q_m=n=0$ 
in eqn.(\ref{pd}) we can recover our result (eqn.(19)
of \cite{cm}) which is applicable for Kerr spacetimes.
\\

{\bf (c) Kerr-Newman spacetime :}
Rotating black hole spacetimes with electric charge $Q_c$ and magnetic
charge $Q$ is described
by Kerr-Newman metric, which is quite important in General Relativity.
Setting $\l=n=0$ in eqn.(\ref{pd}), we  can easily get the LT precession in Kerr-Newman
spacetime. It is thus following (taking $Q_c^2+Q_m^2=Q^2$),

\begin{eqnarray}
\vec{\O}^{KN}_{LT}=\f{a}{\rho^3(\rho^2-2Mr+Q^2)}
\left[\sqrt{\d}(2Mr-Q^2) \cos\t \hat{r}
+(M(2r^2-\rho^2)+rQ^2)\sin\t \hat{\t}\right]
\label{kn}
\end{eqnarray}
In the Kerr-Newman spacetime,
\begin{eqnarray}
 \d=r^2-2Mr+a^2+Q^2  \,\,\,\,
\text{and}\,\,\,\,
\rho^2=r^2+a^2 \cos^2\t
\end{eqnarray}
From the expression (\ref{kn}) of LT precession in Kerr-Newman spacetime
we can see that at the polar plane, {\bf the LT precession
vanishes for the orbit $r=\f{Q^2}{2M}$, though the spacetime
is rotating $(a\neq0)$}. So, if a gyroscope rotates in
the polar orbit of radius $r=\f{Q^2}{2M}$ in this spacetime, the gyroscope
does not experience any frame-dragging effect. So, if any
experiment is performed in future by which we can't 
see any LT precession in that spacetime, it may be happened 
that the specified spacetime is Kerr-Newman black hole
and the gyroscope is rotating at the polar orbit whose 
radius is $r=\f{Q^2}{2M}$. This is a very interesting feature
of the Kerr-Newman geometry. Though the spacetime is rotating
with the angular momentum $J$, the nearby frames are not 
dragged along it. Without this particular orbit the 
LT precession is continued in everywhere in this spacetimes.
\\

{\bf (d) Kerr-de-Sitter spacetimes :}
Kerr-de-Sitter spacetime is more realistic, when
we do not neglect the Cosmological constant parameter 
(though its value is very small, it
may be very useful in some cases, where 
we need to very precise calculation).
Setting $n=Q_c=Q_m=0$, we get the following 
expression for Kerr-de-Sitter spacetime.

\begin{eqnarray}\nonumber
\vec{\O}_{LT}^{KdS}&=& \f{a}{\rho^3(\rho^2-2Mr-
\f{1}{\ell^2}(a^4+a^2r^2-a^4\sin^2\t\cos^2\t))}\left[\sqrt{\d}
(2Mr+\f{1}{\ell^2}\rho^4) \cos\t \hat{r}\right.
\\ && \left.+\sqrt{\x}[M(2r^2-\rho^2)
+\f{r}{\ell^2}\{a^4+a^2r^2-a^4\sin^2\t\cos^2\t-\rho^2(a^2+2r^2)\}])
\sin\t \hat{\t}\right]
\label{kds}
\end{eqnarray}
Where,
\begin{eqnarray}
\d=(r^2+a^2)\left(1-\f{r^2}{\ell^2}\right)-2Mr\,\,\,\,
\text{and}\,\,\,\,
 \x=1+\f{a^2}{\ell^2} \cos^2\t
\end{eqnarray}

\subsection{Non-vanishing Lense-Thirring precession in 
`zero angular momentum'  Pleba\'nski-Demia\'nski spacetimes}

This subsection can be regarded as a special case of
non-accelerating Pleba\'nski-Demia\'nski spacetime in where
we take that PD spacetime is not rotating, we mean the Kerr
parameter $a=0$. In a very recent paper, Hackmann and
L\"ammerzahl shows that Lense-Thirring effect vanishes
(eqn. no (45) of \cite{hl})due to the vanishing Kerr parameter. 
But, we can see easily 
from the eqn.(\ref{pd}) that if $a$ vanishes in PD spacetime,
the LT precession rate will be

\begin{eqnarray}
\vec{\O}^{PD}_{LT}|_{a=0}=\f{n\sqrt{\d|_{a=0}}}{p^3}\hat{r}
\label{pdtn}
\end{eqnarray}
where,
\begin{eqnarray}
\d|_{a=0}=(r^2-n^2)\left(1-\f{1}{\ell^2}(r^2+3n^2)\right)
 &-&2Mr+Q_c^2+Q_m^2-\f{4n^2r^2}{\ell^2} \nonumber
\\
\text{and}, \,\,\,\,\,\,\,\,\,  p^2=r^2&+&n^2
\end{eqnarray}
So, it is not necessary to vanish the LT precession
with vanishing Kerr parameter. The above expression 
reveals that NUT charge $n$ is responsible for the LT
precession in `zero angular momentum' PD spacetimes.
Here, $M$ represents the ``gravitoelectric mass" or `mass' and $n$ represents
the ``gravitomagnetic mass'' or `dual' (or `magnetic') 
mass\cite{lnbl} of this spacetime. It is obvious that the spacetime 
is not invariant under time reversal $t \rightarrow -t$, signifying that 
it must have a sort of `rotational sense' which is
analogous to a magnetic monopole in electrodynamics. One 
is thus led to the conclusion that the source of the 
nonvanishing LT precession is this  ``rotational sense'' 
arising from a nonvanishing NUT charge.
Without the NUT charge, the spacetime is clearly hypersurface 
orthogonal and
frame-dragging effects vanish, as already mentioned in detail
in our previous paper (sec. II of\cite{cm}). 
This {\it `dual' mass} has been investigated in detail in ref. 
\cite{sen2,mp}and it is also referred as an 
{\it `angular momentum monopole} \cite{sen1} in Taub-NUT spacetime.
 This implies that the inertial frame dragging seen here in such
 a spacetime can be identified as a {\it gravitomagnetic} effect.
\\

In that particular paper\cite{hl}, Hackmann and 
L\"ammerzahl investigates the timelike geodesic 
equations in the PD spacetimes. The 
{\it orbital plane precession frequency}
$(\Omega_{\phi}-\Omega_{\theta})$ is computed, 
following the earlier work of ref. \cite{kag, drasco, fh}, 
and a vanishing result ensues. This result is then 
interpreted as a signature for a 
null LT precession in the `zero angular momentum' PD spacetime.

We would like to say that what we have focused 
on in this paper is quite different from the 
`orbital plane precession' considered in \cite{hl}.
 Using a `Copernican' frame, we calculate the precession of a gyroscope
which is moving in an arbitrary integral 
curve (not necessarily geodesic). Within this 
frame, an untorqued gyro in a stationary but not 
static spacetime held fixed
by a support force applied to its center of mass, 
undergoes LT precession. Since the
Copernican frame does not rotate (by construction) relative to the
inertial frames at asymptotic infinity (``fixed stars"), the observed
precession rate in the Copernican frame also gives the precession rate
of the gyro relative to the fixed stars. It is 
thus, more an intrinsic property of the classical 
{\it spin} of the spacetime (as an untorqued gyro
 must necessarily possess), in the sense of a dual 
mass, rather than an {\it orbital} plane precession 
effect for timelike geodesics in a Taub-NUT spacetime. 
\\

In our case, we consider the 
gyroscope equation \cite{ns} in an 
arbitrary integral curve
\begin{equation}
 \nabla_u S=<S,a>u
\end{equation}
where, $a=\nabla_u u$ is the acceleration,
$u$ is the four velocity and $S$ indicates the spacelike classical spin 
four vector $S^{\alpha}=(0,\vec{S})$ of the gyroscope.
 For geodesics $a=0 \Rightarrow \nabla_u S = 0$.

In contrast, Hackmann and L\"ammerzahl \cite{hl} consider the 
behaviour of massive test particles with {\it vanishing spin}
 $S=0$, and compute the orbital plane precession 
rate for such particles, obtaining a vanishing result. 
We are thus led to conclude that because two different 
situations are being considered, there is no 
inconsistency between our results and theirs.

We note that the detailed analyses on LT precession in
Kerr-Taub-NUT, Taub-NUT\cite{taub, nut,msnr} and massless Taub-NUT spacetimes 
has been done in\cite{cm}.

\section{Extremal case}

\subsection{Extremal Pleba\'nski-Demia\'nski Spacetime}
In this section, we would like to describe the LT precession in extremal
Pleba\'nski-Demia\'nski spacetime, whose non-extremal case is already described
in the previous section.
To get the extremal limit in PD spacetimes we should first determine
the radius of the horizons $r_h$ which can be determined by 
setting $\d|_{r=r_h}=0$. We can make a comparison
of coefficients in
\begin{eqnarray}\nonumber
\d&=&-\f{1}{\ell^2} r^4+\left(1-\f{a^2}{\ell^2}-\f{6n^2}{\ell^2}\right)r^2-2Mr
+\left[(a^2-n^2)\left(1-\f{3n^2}{\ell^2}\right)+Q_c^2+Q_m^2\right]\nonumber
\\
&=&-\f{1}{\ell^2}[r^4+(a^2+6n^2-\ell^2)r^2+2M\ell^2r+b]\nonumber
\\
&=&-\f{1}{\ell^2} \Pi_{i=1}^4(r-r_{hi})
\end{eqnarray}
where,
\begin{equation}
b=(a^2-n^2)(3n^2-\ell^2)-\ell^2(Q_c^2+Q_m^2)
\end{equation}
and $r_{hi}$(i = 1, 2, 3, 4) denotes the zeros of $\d$.
From this comparison we can conclude that for the PD-AdS
 (when $\l$ is negative)
black hole, there are two separated positive
horizons at most, and $\d$ is positive 
outside the outer horizon of the PD black hole.
In the same way, we can conclude for the PD-dS
(when $\l$ is positive) black hole that
there are three separated positive horizons at
most, and $\d$ is negative outside the outer
 horizon of the black hole. Both of the
above cases, the when two horizons of the 
PD black hole coincide, the black hole is extremal \cite{li}.

If we consider the extremal PD black hole, 
we have to make a comparison of coefficients in
\begin{eqnarray}
&&\d=(r-x)^2(a_2r^2 + a_1r + a_0)
\\
&=&-\f{1}{\ell^2}[r^4+(a^2+6n^2-\ell^2)r^2+2Mr\ell^2+b]\nonumber
\end{eqnarray}
with $a_0, a_1, a_2$ being real\cite{hl}.
From this comparison we can get the following for PD (``AdS") spacetime,
\begin{eqnarray}
\f{b_{A}}{x^2}-3x^2&=&a^2+6n^2+\ell^2
 \label{li3}
\\
x^3-\f{b_{A}}{x}&=&-M\ell^2
 \label{li4}
\end{eqnarray}
where, $b_A$ represents the value of $b$ at PD (``AdS") spacetimes.
\begin{equation}
 b_A=(a^2-n^2)(3n^2+\ell^2)+\ell^2(Q_c^2+Q_m^2)
\end{equation}

Solving equation (\ref{li3}) for $x$, we get
\begin{equation}
x=\sqrt{\f{1}{6}\left[-(\ell^2+a^2+6n^2)+\sqrt{(\ell^2+a^2+6n^2)^2+12b_A}\right]}
\label{li5}
\end{equation}

Similarly, we can obtain for PD(``dS") black hole ,
\begin{eqnarray}
\f{b}{x^2}-
3x^2&=&a^2+6n^2-\ell^2
\label{li1}
\\
x^3-\f{b}{x}&=&M\ell^2
\label{li2}
\end{eqnarray}

In these equation $x$ is positive and related to the coincided
horizon of the extremal PD (for ``dS") black hole.

Solving equation (\ref{li1}) for $x$, we get

\begin{eqnarray}
x_+&=&\sqrt{\f{1}{6}\left[\ell^2-a^2-6n^2+\sqrt{(\ell^2-a^2-6n^2)^2+12b}\right]}
\\
x_-&=& \sqrt{\f{1}{6}\left[\ell^2-a^2-6n^2-\sqrt{(\ell^2-a^2-6n^2)^2+12b}\right]}
\end{eqnarray}

where, $x_+$ and $x_-$ indicate the outer horizon and inner horizon, respectively.
This can be seen by calculating
\begin{equation}
 \f{d^2 \d}{dr^2}|_{r=x_+}=-\f{2}{\ell^2}\sqrt{(\ell^2-a^2-6n^2)^2+12b}
\end{equation}
and
\begin{equation}
 \f{d^2 \d}{dr^2}|_{r=x_-}=\f{2}{\ell^2}\sqrt{(\ell^2-a^2-6n^2)^2+12b}
\end{equation}
For the PD(``dS") black hole, on the outer extremal horizon,
$\f{d\d}{dr}=0$ and $\f{d^2 \d}{dr^2}<0$ and on the inner
extremal horizon $\f{d^2 \d}{dr^2}>0$.
Now, we can solve $M$ and $a$ from the two equations(\ref{li1}, \ref{li2}).

\begin{eqnarray}
M &=&\f{x\left[x^4+2x^2(3n^2-\ell^2)+(3n^2-\ell^2)(7n^2-\ell^2)+\ell^2(Q_c^2+Q_m^2)\right]}
{\ell^2(\ell^2+x^2-3n^2)}
\label{me}
\\
a_e^2 &=&\f{3x^4+(6n^2-\ell^2)x^2+n^2(3n^2-\ell^2)+\ell^2(Q_c^2+Q_m^2)}
{(3n^2-\ell^2-x^2)}
\label{ae}
\end{eqnarray}
From the above values of $a_e^2$ and $M$, we get
\begin{equation}
a_e^2=-M\ell^2\f{\left[3x^4+(6n^2-\ell^2)x^2+n^2(3n^2-\ell^2)+\ell^2(Q_c^2+Q_m^2)\right]}
{x\left[x^4+2x^2(3n^2-\ell^2)+(3n^2-\ell^2)(7n^2-\ell^2)+\ell^2(Q_c^2+Q_m^2)\right]}
\label{ame}
\end{equation}
The ranges of $x$ and $a_e$ are determined from the following expressions
\begin{eqnarray}
x^2&<&\left(\f{\ell^2+\ell \sqrt{\ell^2-12Q^2}}{6}-n^2 \right)
\label{xe}
\\
0<a_e^2&<&\left[(7\ell^2-24n^2)-\sqrt{(7\ell^2-24n^2)^2-(\ell^{4}-12\ell^2(Q_c^2+Q_m^2))}\right]
\label{a2e}
\end{eqnarray}

Due to the presence of Cosmological constant, there exist four 
roots of $x$ in eqn. (\ref{ae}). 
When Cosmological constant $\f{1}{l^2}\rightarrow 0$,
eqn.(\ref{me}) and eqn.(\ref{ae}) reduces to
\begin{equation}
 x=M
\end{equation}
and,
\begin{equation}
 a_e^2=x^2+n^2-Q_c^2-Q_m^2
\end{equation}
or,
\begin{equation}
 a_e^2=M^2+n^2-Q_c^2-Q_m^2
\end{equation}
respectively.

Now, the line element of extremal PD spacetimes can be written as,

\begin{eqnarray}
ds^2&=&-\f{\d_{e}}{p_e^2}(dt-A_ed\phi)^2+\f{p_e^2}{\d_{e}}dr^2 
+\f{p_e^2}{\x_{e}}d\t^2+\f{\x_{e}}{p_e^2}\sin^2\t(a_edt-B_ed\phi)^2
\label{lnelmntx}
\end{eqnarray}

and, the final LT precession rate in extremal PD spacetime is,

\begin{eqnarray}\nonumber
\vec{\O}_{LT}^{ePD}&=&\f{\sqrt{\d_{e}}}{p_e}\left[\f{a_e(\x_{e} \cos\t
+(2n-a_e\cos\t)\f{a_e}{\ell^2} \sin^2\t)}{\d_{e}-a_e^2\x_{e} \sin^2\t}
-\f{a_e \cos\t-n}{p_e^2}\right]\hat{r}
\\
&+&\f{\sqrt{\x_{e}}}{p_e}a_e \sin\t \left[\f{r-M-\f{r}{\ell^2}(a_e^2+2r^2+6n^2)}
{\d_{e}-a_e^2\x_{e} \sin^2\t}-\f{r}{p_e^2}\right]\hat{\t}
\label{pdx}
\end{eqnarray}
where,
\begin{eqnarray}\nonumber
\d_{e}=-\f{1}{\ell^2}\left(r-x \right)^2 \left(r^2+2rx+\f{b}{x^2}\right) , \,\,\,\,
\x_{e}=1+\f{a_e^2}{\ell^2} \cos^2\t-\f{4a_en}{\ell^2} \cos\t \nonumber
\\
p_e=r^2+(n-a_e\cos\t)^2 , \,\,\,\, A_e=a_e \sin^2\t+2n \cos\t ,
\,\,\,\,
B_e=r^2+a_e^2+n^2
\label{deffn1}
\end{eqnarray}
and,  the value of $a_e$ is determined from the eqn.(\ref{ae}) and the range
of $x$ and $a_e$ (`$e$' stands for the {\it extremal} case) 
are determined from the eqn. (\ref{xe}) and (\ref{a2e}),
respectively. It could be noted that, for the extremal PD(``dS"),
spacetimes, {\it there are upper limiting values for angular momentum
and extremal horizon of the black hole}.  
Substituting all the above mentioned values and ranges in eqn.(\ref{pdx}),
we get the exact LT precession rate in extremal PD spacetimes.  


\subsection{Extremal Kerr Spacetime}

Substituting $a_e=M$ in  equation (19) of \cite{cm}) or substituting $a=M$
and $\l=Q_c=Q_m=n=0$ in equation (\ref{pdx}) we can easily get the LT
precession rate in extremal Kerr spacetime.
\begin{eqnarray}
\vec{\O}_{LT}^{eK}= \f{M^2[2r(r-M)\cos\t \hat{r}+(r^2-M^2\cos^2\t)
\sin\t \hat{\t}]}{(r^2+M^2\cos^2\t)^{\f{3}{2}}(r^2-2Mr+M^2\cos^2\t)}
 ~\label{maLT}
\end{eqnarray}
The above result is also coming from the eqn.(\ref{pdx}), the general LT precession
rate for extremal PD spacetime.
Case I: On the polar region i.e. $\theta =0$ the $\Omega_{LT}$ becomes
\begin{eqnarray}
\Omega_{LT}^{eK}=\f{2M^{2}r}{(r^{2}+M^{2})^{3/2}(r-M)} ~\label{elt}
\end{eqnarray}

Case II: On the equator i.e. $\theta =\pi/2$, the $\Omega_{LT}$ becomes
\begin{eqnarray}
\Omega_{LT}^{eK}=\f{M^{2}}{r^{2}(r-2M)} ~\label{eltc}
\end{eqnarray}
It could be easily seen that $\O_{LT}$ is diverges at $r=M$. Since $r=M$
is the only direct ISCO in extremal Kerr geometry which coincides with
the principal null geodesic generator of the horizon\cite{pp} and it is also the
radius of the horizon which is a null surface. The general LT precession
formula is derived only considering that the observer is rest in a
timelike Killing vector field. We have not incorporate the LT effect
for any null geodesics. So, our formula is valid for $r>M$ (outside the horizon).


\subsection{Extremal Kerr-Newman Spacetime}
Substituting  $a^2=M^{2}-(Q_{c}^{2}+Q_m^2)=M^{2}-Q^{2}$
and $\l=n=0$ in equation (\ref{kn}),
we can obtain the following LT precession rate for extremal KN blackhole

\begin{eqnarray}
\vec{\O}^{eKN}_{LT}&=& \f{\sqrt{M^2-Q^2}}{\rho^3(\rho^2-2Mr+Q^2)}
\left[(r-M)(2Mr-Q^2)\cos\t \hat{r}
+(M(2r^2-\rho^2)+rQ^2)\sin\t \hat{\t}\right]
\label{ekn}
\end{eqnarray}

where
\begin{equation}
 \rho^2=r^2+(M^2-Q^2)\cos^2\t
\end{equation}
From the above expression(eqn. \ref{ekn}), we can make 
a similar comment again, like the extremal Kerr-Newman black hole
that the gyroscope which is rotating in a polar orbit
of radius $r=\f{Q^2}{2M}$, cannot experience any frame-dragging effect.
Apparently, it seems that this argument is also true
for the gyroscope which is rotating at $r=M$ orbit. 
But, this is not true. Because, 
this is the horizon of extremal Kerr-Newman spacetime. So, $r=M$ is a
null surface. The general formula which we
have considered in our whole paper, is valid only 
in timelike spacetimes (outside the horizon), not any 
null or spacelike regions.


\subsection{Extremal Kerr-de Sitter Spacetime}

Extremal Kerr-de Sitter spacetime is interesting because it involves the
Cosmological constant. Setting $n=Q_c=Q_m=0$ in eqn.(\ref{pd}), we can
find the following
expression of LT precession at extremal Kerr-de Sitter spacetimes

\begin{eqnarray}\nonumber
\vec{\O}_{LT}^{eKdS}&=& \f{a_e}{\rho^3(\rho^2-2Mr-
\f{1}{\ell^2}(a_e^4+a_e^2r^2-a_e^4\sin^2\t\cos^2\t))}\left[\sqrt{\d_e}(2Mr+\f{\rho^4}{\ell^2})
\cos\t \hat{r}\right. \\ && \left.+\sqrt{\x_e}[M(2r^2-\rho^2)+
\f{r}{\ell^2}\{a_e^4+a_e^2r^2-a_e^4\sin^2\t\cos^2\t-\rho^2(a_e^2+2r^2)\}])\sin\t \hat{\t}\right]
\label{kds}
\end{eqnarray}

where,
\begin{eqnarray}
&& \rho^2=r^2+a_e^2 \cos^2\t \,\,\, , \,\,\,\, \x_{e}=1+\f{a_e^2}{\ell^2} \cos^2\t
\\
\text{and} && \nonumber
\\
&& \d_{e}=-\f{1}{\ell^2}(r-x)^2\left(r^2+2xr-\f{a^2\ell^2}{x^2}\right) \nonumber 
 \label{deffne}
\end{eqnarray}

Using equations (\ref{ae}, \ref{me}), we obtain the value for $a$ and $M$ are
\begin{eqnarray}
a_e^2 &=&\f{(\ell^2-3x^2)x^2}{(\ell^2+x^2)}   \label{ae1}\nonumber\\
M &=&\f{x(\ell^{2}-x^{2})^{2}}{\ell^2(\ell^2+x^2)}
\label{me1}
\end{eqnarray}
where the horizons are  at
\begin{eqnarray}\nonumber
x_+&=&\sqrt{\f{1}{6}\left[\ell^2-a^2+\sqrt{(\ell^2-a^2)^2-12a^{2}\ell^{2}}\right]}
\\
x_-&=& \sqrt{\f{1}{6}\left[\ell^2-a^2-\sqrt{(\ell^2-a^2)^2-12a^{2}\ell^{2}}\right]}
\end{eqnarray}
where, $x_+$ and $x_-$ indicate the outer horizon and inner horizon.
The range of $a_e^{2}$ and $x$ are the following.
\begin{equation}
0<a_e^{2}<(7-4\sqrt{3})\ell^{2} \,\,\,\,
\text{and} \,\,\,\, x^2<\ell^2/3 
\end{equation}
which is already discussed in \cite{li}. Substituting all the above values 
in eqn.(\ref{kds}) and taking the ranges of $a_e$ and $x$, we get the 
exact LT precession rate in extremal Kerr-dS spacetimes.


\subsection{Extremal Kerr-Taub-NUT spacetime}

To derive the extremal limit in Kerr-Taub-NUT spacetime we set,
\begin{equation}
 \d=r^2-2Mr+a^2-n^2=0
\end{equation}
Solving for $r$, we get two horizons which are located at $r_{\pm}=M\pm\sqrt{M^2+n^2-a^2}$.
So, we get the extremal condition $(r_+=r_-)$ for Kerr-Taub-NUT spacetimes
is $a_e^2=M^2+n^2$.
If we set $Q_c=Q_m=\l=0$ and $M^{2}+n^{2}=a_e^{2}$ in eqn. (\ref{pd}), we get
the following exact LT precession rate at extremal Kerr-Taub-NUT spacetime.

\begin{eqnarray}\nonumber
\vec{\O}^{eKTN}_{LT}&=&\f{(r-M)}{p}\left[\f{\sqrt{M^2+n^2}\cos\t}{p^2-2Mr-n^2}
-\f{\sqrt{M^2+n^2}\cos\t-n}{p^2}\right]\hat{r}
\\
&+&\f{\sqrt{M^2+n^2}\sin\t}{p} \left[\f{r-M}{p^2-2Mr-n^2}
-\f{r}{p^2}\right]\hat{\t}
\label{ekt}
\end{eqnarray}

Where
$p^2=r^2+(n\mp\sqrt{M^2+n^2}\cos\t)^2$.

\section{Discussions}
In this work we have explicitly derived the Lense
Thirring precession frequencies for extremal and non-
extremal Pleba\'nski-Demia\'nski(PD) spacetime. The
PD family of solutions are the solutions of Einstein
field equations which contain a number of well
known black hole solutions. Among them including
Schwarzschild, Schwarzschild-de-Sitter, Kerr, Kerr-
ads, Kerr-Newman, Kerr-Taub-NUT etc. We observe
that LT precession frequency strongly depends upon
the different parameters like mass $M$, spin $a$, Cosmological constant $\l$,
NUT charge $n$, electric charge $Q_c$ and magnetic charge $Q_m$. An
interesting point that LT precession occurs solely due to the ``dual
mass". This ``dual mass" is equivalent to angular momentum
 monopole $(n)$ of NUT spacetime. For our
completeness we have also deduced the LT precession for extremal
PD spacetime and also for others axisymmetric PD-like spacetimes.
To get the LT precession rate in extremal Kerr-Taub-NUT-de-Sitter spacetime, 
the basic procedure is same as the PD spacetimes with the additional 
requirement is $Q_c^2+Q_m^2=0$ in equations (\ref{me}) and (\ref{ae}) 
and the range of $x^2<(\f{\ell^2}{3}-n^2)$ and
$a_e^2$ has a range of $0<a_e^2<(7\ell^2-24n^2)-4\sqrt{3(\ell^2-3n^2)(\ell^2-4n^2)}$.
As there is no any valid
extremal condition at NUT spacetime (with $M$ or without $M$),
we could not get any {\it real} LT precession rate due to frame-dragging 
effect. Since, the direct ISCO coincides with the principal null
geodesic generator\cite{pp} in extremal Kerr spacetimes
and Kerr-Newman spacetimes, we
are unable to discuss the LT precession at that particular geodesic.
So this formula is not valid
for the domain of $r \leq M$ for the extremal Kerr
and Kerr-Newman spacetimes. Here,
our formula is valid only for $r>M$. In general, the general
formula for LT precession in stationary
spacetime is valid only outside the horizon, as
the observer is in timelike Killing vector field. The formula is
not valid on the horizon and inside the horizon.
We will discuss about this problem in near future.

\textbf{Acknowledgments :} We would like to thank Prof. Parthasarathi Majumdar for his
encouragement and guidance, besides the long blackboard discussions on this topic.

\end{document}